# Optoacoustic Signal-Based Underwater Node Localization Technique: Overcoming GPS Limitations without AUV Requirements

Maliha Tasnim[1] and Rashed Hasan Ratul[2]
[1]Department of Mechanical and Production Engineering
[2]Department of Electrical and Electronic Engineering
Islamic University of Technology (IUT)
Dhaka, Bangladesh
{malihatasnim, rashedhasan}@iut-dhaka.edu

*Abstract*— Underwater sensor networks are anticipated to facilitate numerous commercial and military applications. Moreover, precise self-localization in practical underwater scenario is a crucial challenge in sensor networks because of the complexity of deploying sensor points in specific locations. The Global Positioning System (GPS) is inappropriate for underwater localization because saline water may severely attenuate signal, limiting penetration capacity to barely a few meters. Hence, the most promising alternative to wireless radio wave transmissions for underwater networks is regarded as acoustic communication. In order to establish an underwater localization model, traditional techniques that essentially include surface gateways or intermediate anchor nodes create logistical challenges and security hazards when the surface access points are deployed. This paper proposes a unique localization method that employs optoacoustic signals to remotely localize underwater wireless sensor networks in order to address these concerns. In our model, the GPS signals are transmitted from the air to the underwater node through a mobile beacon that employs optoacoustic techniques to generate a short-term isotropic acoustic signal source. Underwater nodes with omnidirectional receivers can detect their location in random and realistic environments by comparing signal levels from two equivalent plasmas, also referred as acoustic source nodes. Finally, the software simulation results in comparison to the proposed theoretical model highlight the feasibility of our approach. In comparison to complicated traditional node localization approach, our simplified approach is anticipated to achieve better accuracy with only a single off-water controlled node, while avoiding the additional requirement for surface anchor nodes.

## I. INTRODUCTION

The advancement of underwater communication and technology has opened up new opportunities for the analysis of the oceans and the aquatic environment. Underwater wireless sensor networks are gaining prominence due to their potential to advance ocean research and meet the requirements of a variety of undersea usage, including gathering oceanographic data, safety systems for natural catastrophes, search and rescue, assisted navigation, coastal patrol, security surveillance, and many others [1]. For a holistic understanding of the acquired data from the Underwater Nodes (UWNs), it is essential to know where the nodes are deployed. The usage of the commonly used GPS is only compatible with surface nodes since RF communications are severely degraded underwater [2]. Since GPS cannot be implemented to localize UWN, an alternative and feasible approach must be utilized to facilitate localization-related information exchange across submerging UWNs and surface nodes.

Rather than employing GPS-based localization, the conventional and existing approaches use relative coordinate methods based on submerged anchor nodes. In particular, it has been commonly considered that a surface gateway that would function as an anchor node or plasma will be deployed to function as an interconnection, employing radio signals for air-based communication and acoustic signals for underwater communication [3]. The Autonomous Underwater Vehicle (AUV) receives GPS signals when floating, descends to a predetermined depth, and moves through a specific track among sensor nodes to collect data from the entire field [4]. However, there are a number of significant drawbacks to having a floating network based node, including the logistical constraint and additional time required for dive and rise technique that makes deployment challenging. Furthermore, such implementation might put the underwater network at risk for security breaches and the additional expenses and equipment requirements make operating the mobile underwater networks more complicated. Avoiding the AUV and surface gateway necessitate the improvements of a localization method utilizing cross-medium communication techniques. Because of the considerable attenuation that occurs inside the water medium, the majority of wireless signals cannot offer meaningful underwater propagation. Optoacoustic signals have therefore been the most appropriate approach of communication in the aquatic environment [5].

In this paper, a reasonable alternative for performing UWN localization without using surface-based reference nodes or AUV is proposed. The theory is to use optoacoustic signals to establish communication links between different mediums [3]. A potential alternative to the cross-medium air-water communication barrier is to integrate the optoacoustic effect [6]. The optoacoustic phenomenon was initially discovered and identified in 1881 by Alexander Graham Bell [7]. He observed that an acoustic wave is produced when higher intensity level light strikes a liquid substance like water. The optoacoustic radiation transformation mechanism can be separated into two different strategies: linear mechanism and nonlinear mechanism. The characteristics of aqueous medium remain unchanged in the linear mechanism. Conversely, a nonlinear optoacoustic method changes the aqueous medium's physical characteristics, specifically by turning water into vapor and inducing cavitation bubbles [8]. The nonlinear optoacoustic method produces a higher Sound Pressure Level (SPL) over its linear version, making it appropriate for communicating with underwater sensors located far below the surface. According to the simulation results in [8], the SPL of a certain linear optoacoustic procedure can produce equal to 140 dB re 1 Pa. Conversely, well over than 210 dB re-Pa at 1 m have been documented in [9], when it comes to the SPL of a nonlinear way of an optoacoustic effect.

In order to address the issues of GPS limitations, 3-D spatial orientation, random orientation of underwater nodes, limited message transmission, and minimal computational requirements, an innovative localization strategy has been presented for underwater sensor networks. The suggested technique entails sending a localization information block comprising GPS coordinates from the air to the UWN using a laser beam concentrated in water. By calculating distances from acoustic transmitters utilizing Received Signal Strength (RSS), the UWN calculates its own GPS coordinates. This approach is anticipated to address physical layer limitations, and simulation findings demonstrate that the proposed technique achieves relatively better accuracy compared to that of conventional approaches that require the use of underwater anchor nodes. The feasibility of this strategy was verified through simulation employing experimentally gathered data obtained from [10].

The remaining part of the paper proceeds as follows: Section 2 presents the related works. Section 3 examines the method of producing an optoacoustic signal. The methodology and the proposed underwater node localization technique is covered in Section 4 and 5, respectively. Section 6 discusses the performance parameters. Finally, Section 7 draws the concluding remarks.

## II. RELATED WORKS

Numerous techniques have been presented in recent years for underwater node localization. Most published localization techniques make use of acoustic signals since they are the better choice for underwater environments. Additionally, in underwater installations, the acoustic signal produced by the optoacoustic process typically has a longer range of propagation than visible light.

Nguyen et al. proposed RSS-based localization strategies where anchors are randomly dropped to determine the location of an unknown node [11]. To achieve low error and high precision, however, this approach requires preprocessed data generated via an iterative localization procedure. The Time Difference of Arrival (TDoA)-based localization model proposed by Liu et al. uses multiple surface beacons to conserve energy, but this method may come with added expenses, logistical complexities, and security vulnerabilities [12]. Luo et al. present a novel method for underwater localization that utilizes directional signals transmitted from an AUV [13]. This approach employs directional beacons as opposed to conventional omni-directional techniques, resulting in more precise and effective sensor localization through simple calculations. However, they didn't extend their approach to the three-dimensional problem by taking into account water depth, which is a crucial factor in achieving accurate localization. Zhou et al. proposed a distributed localization technique that involves the dynamic integration of a recursive location estimation method and a 3D Euclidean linear distance estimating method. The localization operation is carried out through the mutual processes of anchoring node localization and random underwater node localization. However, only surface buoys are initially installed with information about their locations, which are obtained through standard GPS or other means. The method requires at least four buoys to function, which serve as the network's satellites [14]. The process of localizing nodes was introduced in [15], using an AUV and the authors discussed the associated trade-offs, including the proportion of localized sensor nodes and the accuracy of localization. Two localization methods were explored: triangulation and bounding-box. The paper proposed a straightforward underwater GPS system that employs a single AUV, which can be used to construct an underwater wireless sensor network. To strike a balance between performance and cost, the paper also suggested using a large number of sensors. However, the effectiveness of AUV-aided localization in mobile scenarios will need to be investigated.

Lee et al. introduce a novel localization scheme that uses a pressure sensor and a mobile beacon [16]. The proposed range-free localization approach involves the use of a mobile beacon to estimate two possible locations of a wireless sensor node based on mathematical and geometric calculations and then select the final location. However, further investigation into route planning for the mobile beacon is needed as the path it takes can significantly impact the accuracy of localization. Teymorian et al.'s study focuses on the localization of 3D underwater wireless networks. The task of network localization involves identifying the precise position of all the node in the setup, using the positions of certain nodes and internode distances [17]. The article explains the conditions that allow for the transformation of the three-dimensional localization method into a two-dimensional version and introduces a distributed localization agenda that leverages projections to determine the underwater sensor positions. Cheng et al. suggest their submerged positioning scheme incorporating beacon signals via four anchors transmitted in consecutive time slots, the TDoA has been implemented for underwater localization [18]. However, this method is limited only for the stationary underwater acoustic sensor networks. The authors of [19] suggest an alternative solution involving the use of movable anchor nodes that can obtain GPS coordinates while above water and occasionally broadcast their current location while going beneath the surface.

Mahmud et al. propose a new method for localizing underwater networks using optoacoustic signals, which eliminates the need for floating gateways [10]. The approach creates an underwater acoustic transmitter through the optoacoustic effect to transmit GPS coordinates from air to the UWN. Future research and systematic investigations may explore ways to improve the accuracy and efficiency of the optoacoustic approach and investigate its performance in more complex underwater environments. However, it is common for conventional underwater localization models to use either surface nodes or AUVs to facilitate the transmission of GPS signals to UWNs. The aim of this paper is to create a localization technique for decentralized and randomized UWNs that does not rely on surface buoy or AUVs. Additionally, only one airborne node is sufficient in our proposed approach to accurately locate the position of UWNs that would open up the possibilities of minimizing additional expenses and logistical requirements. We capitalize on the primary benefits of alternate cross-medium transmission of acoustic signals, which has a significantly higher underwater communication range.

## III. Nonlinear Optoacoustic Generaion

A powerful laser transmitter is employed in the optoacoustic method to produce underwater acoustic waves from a distant aerial position. Energy transfer triggers a certain change in the overall physical state of the medium throughout the nonlinear optoacoustic phase, which causes optical breakdown. The process of generating plasma is triggered when the laser parameters exceed a predetermined breakdown threshold value, determined by the specific settings of the laser. The minimum intensity required to induce optical breakdown in water ranges from $10^{11}$ W/cm$^2$ over a few nanosecond pulses to $10^{13}$ W/cm$^2$ for 100 femtosecond pulses [20]. The acoustic signal is produced by the collapsed shockwave and consecutive cavitation bubbles expansion-collapse shockwaves, both of which are caused by the generation of plasma during the optical breakdown. Our proposed localization method replaces conventional submerged transducers with this technology to convey coordinate signals from a remote airborne site. The duration and beam-width of the acoustic pulses are influenced by the shape and size of the acoustic source generated by the laser [6]. Isotropic pressure in all directions is required to provide precise distance estimation for UWN localization. As a result, perhaps a more spherical plasma structure is required, which can be accomplished by altering the laser's parameters [10]. Authors in [21] developed an efficient modulation technique for optoacoustic communication to optimize the overall throughput and power efficiency using a single laser beam. In this paper, we have utilized the experimental findings obtained from [10], where a laboratory experiment was carried out to demonstrate the generation of nonlinear optoacoustic signals.

## IV. Methodology

The localization accuracy is very crucial for an UWN. In this optoacoustic system, a laser transmitter is directed from air to water and focused beneath the surface. As the mediums differ, the speed at which the laser propagates also varies and is not identical to the velocity of the acoustic signal. Thus, the most practical technique used for distance measurement is the RSS-based method as all the other distance measuring techniques such as Angle of Arrival (AoA), Time of Arrival (ToA), and TDoA are impractical in this context. The aerial node analyzes the plasma's location employing its GPS coordinates. The underwater plasma produces acoustic signals that are emitted equally in all directions, and the UWN determines proximity by analyzing the measured RSS of these signals. Initial signal strength or the SPL is considered to be determined by the corresponding airborne node for its specific laser specifications and is transmitted inside the signal packet to the unlocalized UWN. Finally, the acoustic transmitter's coordinates—which correspond to the location of the acoustic sources—are transferred to the UWN. To compute the Transmission Loss (TL) accurately, the UWN employs the average Sound Intensity Level (SIL) from the signal blocks it receives, which is used to calibrate and calculate the TL,

$$TL = SPL - SIL \quad (1)$$

However, in the water, acoustic signal transmission is decreased because of two processes: spreading and absorption [22]. The entire value of TL is provided by,

$$TL = 10 \cdot k \cdot \log R + (\alpha \cdot R \cdot 10^{-3}) \quad (2)$$

In this equation, $k$ represents the spreading factor which is set to be 2 for spherical spreading that is being considered in our proposed model. $R$ indicates the distance between the receiver UWN and the plasma, and $\alpha$ signifies the absorption co-efficient.

In Equation (2), spreading factor is shown in the first half and absorption loss is shown in the second half. The dispersion of sound waves due to various forms of inhomogeneity is another reason why sound strength decreases with distance. The formula developed by Thorp is used to calculate the absorption coefficient, $\alpha$ [23],

$$\alpha = \frac{0.11 f^2}{1+f^2} + \frac{44 f^2}{4100+f^2} + 2.75 \cdot 10^{-4} f^2 + 0.003 \quad (3)$$

Here, the frequency $f$ is denoted in kHz, and the absorption coefficient here is denoted as $\alpha$ in decibels per kilometer (dB/km).

Employing the standardized Lambert $W$ function to invert the $TL$ from (2) and the Halley technique to determine $R$ [24],

$$R = \frac{20000 \cdot W\left(\left(\frac{\ln(10)}{20000}\right) a e^{\left(\ln\frac{(10)}{20}\right) \cdot TL}\right)}{a \cdot \ln(10)} \quad (4)$$

It is the formula for measuring the distance from the plasma. The Lambert $W$ function is much more feasible and efficient in calculating an accurate distance.

## V. UWN Localization Using Optoacoustic Signals

The proposed localization technique for UWN needs to be installed with a pressure sensing module and a receiver. The depth from the water's surface will be recorded using a pressure sensor and the receiver would analyze the direction of any received signal from which it is transmitted to the UWNs. In this case, just one specific position of the aerial node is necessary to concentrate the laser transmitter into the aqueous medium and broadcast the localization information blocks. The UWNs that have not been localized receive the localization message blocks from every specific location of the aerial unit that proceeds along the x direction only.

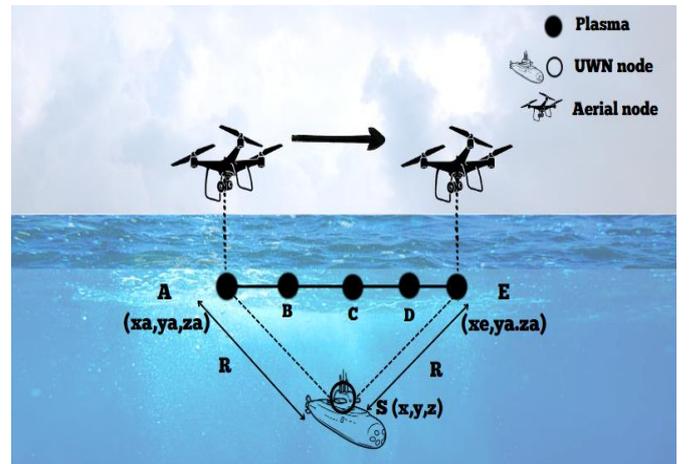

**Fig. 1.** Underwater node localization employing optoacoustic signals by an aerial node.

Fig. 1 depicts the movement of an aerial node in a straight path, while continually creating acoustic sources inside the water medium at a constant depth. Hence, this aerial node generates acoustic sources at point A, B, C, D, and E respectively and transmit the optoacoustic signals. This horizontally moving aerial node creates acoustic sources only varying its x-coordinate while y and z-coordinate will remain fixed as illustrated in Fig. 1. Any unknown UWN, $S$ that falls within the communication range of the acoustic sources, can receive a sequence of message blocks from them. Consequently, the unknown node $S$ can detect the acoustic sources at that moments A, B, C, D, E and so on, which are designated as acoustic sources. According to Fig.1, A and E correspond to the necessary instances here, as we consider that the node $S$ perceives the same pressure from these two acoustic sources. Interestingly, it is not necessary for node $S$ to store the information received from all other acoustic sources. Rather, solely the information of acoustic sources i.e., A and E are sufficient to pinpoint the location of $S$ since it receives the same amount of pressure from the points A and E as mentioned earlier. If the location of unknown node $S$ is within range of RSS-based distance $R$, as illustrated in Fig. 1, it would be capable of recording the coordinates $(x_i, y_i)$ of any acoustic source. In this scenario, $S$ must be able to detect the acoustic source when it transmits at A $(x_a, y_a)$ and then identify it again when it transmits at E $(x_e, y_a)$. Based on this information, $S$ can then estimate its own position $(x, y)$ by means of certain simplified calculations:

$$x = \frac{x_a + x_e}{2} \quad (5)$$

$$d = \sqrt{R^2 - (z - z_a)^2} \quad (6)$$

$$y = y_a \pm \frac{1}{2}\sqrt{(2d)^2 - (x_a - x_e)^2} \quad (7)$$

It should be noted that the distance $R$ measured from the RSS-method for localization is considered as real-world three-dimensional distance. However, we can convert this distance to two-dimensional counterpart which is shown in Equ. (6) and it is denoted by $d$. The significance of using both sign ($\pm$) in Equation (7) depends on the direction of arrival of the signal from the plasma or acoustic source detected by the directional receiver on $S$. Additionally, whether to apply ($\pm$) can also be detected by co-ordinate information. As a result, the UWN's position can be determined by obtaining localization message blocks that correspond to multiple spots of the airborne node as it moves along the water's surface.

VI. RESULT AND DISCUSSION

Matlab simulation has been employed to evaluate the proposed underwater localization approach. Our simulation setup for practical, realistic, and randomized UWN localization can be illustrated in Fig. 2, where acoustic sources are produced in several collinear locations along the horizontal axis with respect to the water surface level for transmitting the localization message blocks. For simplicity, only two acoustic sources or plasmas have been illustrated in Fig. 2 and these acoustic plasma sources are considered to be equivalent sources in terms of pressure received at the specific UWN location. However, there will be several numbers of plasma generated at a constant depth underwater, but at a time, only two equivalent plasma sources will be considered by any random UWN. For accurate localization, the UWN must receive an equivalent matching pressure level from two different plasmas at the same instance.

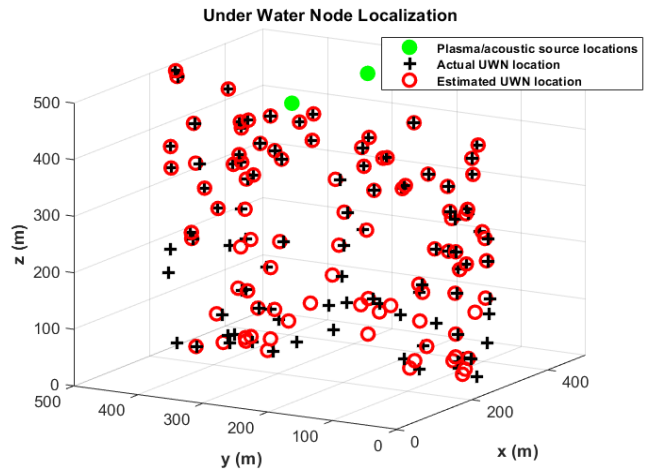

**Fig. 2.** An illustration of multiple UWN deployment environment and their anticipated locations.

For the transmission of acoustic signals underwater, Additive White Gaussian Noise (AWGN) channels are taken into consideration. During the simulation, 100-randomly oriented UWNs are considered in a 500×500×500 cubic meter three-dimensional area. Hence, our proposed technique is used to estimate the UWN position. The Root Mean Square Error (RMSE) quantifies the accuracy of our proposed approach. The RMSE is expressed by [25],

$$RMSE = \sqrt{\frac{\sum_{i=1}^{N}(x_i - \tilde{x}_i)^2 + (y_i - \tilde{y}_i)^2 + (z_i - \tilde{z}_i)^2}{N}} \quad (8)$$

In this case, the actual UWN position is $(x_i, y_i, z_i)$, the anticipated location is $(\tilde{x}_i, \tilde{y}_i, \tilde{z}_i)$, and $N$ represents the total quantity of unlocalized UWN. Here, Fig. 3 shows the localization error for such a randomized UWN localization arrangement at different Signal to Noise Ratio (SNR) levels.

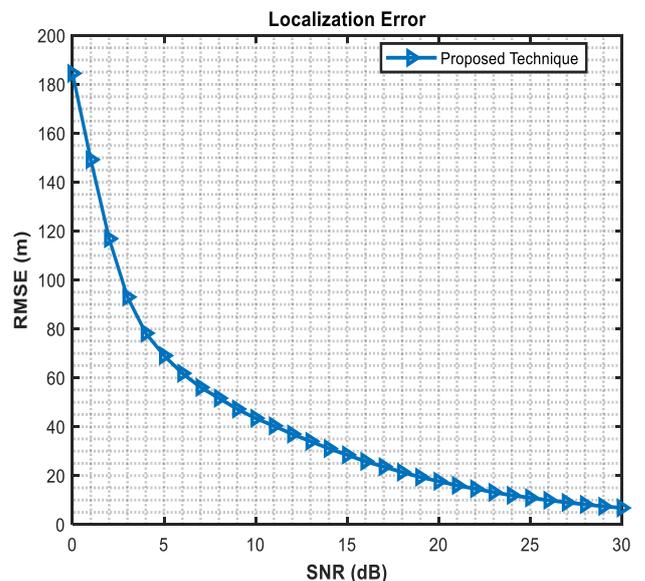

**Fig. 3.** RMSE vs. SNR for UWN position.

Contrary to conventional underwater localization methods utilizing surface buoys or AUVs, our proposed method showcases a significant reduction in localization error, as depicted in Fig. 4. Our proposed technique performs much better at lower SNR levels and sharp drop in the RMSE is observed when SNR is increased.

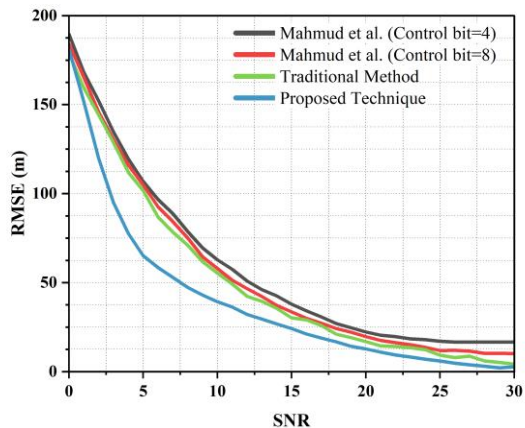

**Fig. 4.** Comparison of RMSE vs. SNR for UWN localization [10].

## VII. CONCLUSION

This paper introduces a novel approach for underwater node localization and GPS coordinate transmission through optoacoustic signals. To address the randomized nodes in the ocean, we came up with an innovative solution. Simulation results have validated the feasibility of our method. The validation results demonstrate that our strategy surpasses traditional methods, particularly those reliant on additional AUVs, in terms of accuracy across various signal-to-noise ratio (SNR) ranges. In conclusion, the optoacoustic-based methodology presented in this paper can potentially play a significant role in future research findings related to underwater localization and communication. In addition, it has the potential to provide a practical and cost-effective alternative for relatively accurate and efficient localization of underwater nodes, thereby eliminating the necessity of logistically challenging and costly equipment such as AUVs or underwater anchor nodes.

## ACKNOWLEDGMENT

We would like to express our sincere gratitude to Mr. Muntasir Mahmud, a researcher from the University of Maryland, Baltimore County, for sharing his knowledge and expertise on underwater optoacoustic signals to successfully conduct this research.